\documentclass[aps,prc,showpacs,twocolumn,preprintnumbers,floatfix,
showkeys,tightenlines]{revtex4}
\usepackage{amsmath,amssymb,amsfonts}
\usepackage{graphicx}
\usepackage{color}
\allowdisplaybreaks

\newcommand{\be}{\begin{equation}}
\newcommand{\ee}{\end{equation}}
\newcommand{\bea}{\begin{eqnarray}}
\newcommand{\eea}{\end{eqnarray}}
\newcommand{\bm}{\bibitem}

\begin{document}



\title{ Temperature dependence of symmetry energy of finite nuclei}

\author{J. N. \surname{De}}
\email{jn.de@saha.ac.in}
\author{S. K. \surname{Samaddar}}
\email{santosh.samaddar@saha.ac.in}
\affiliation{
Saha Institute of Nuclear Physics, 1/AF Bidhannagar, Kolkata
{\sl 700064}, India} 


\begin{abstract} 
The temperature dependence of the symmetry energy and the symmetry free 
energy coefficients of atomic nuclei is investigated in a finite temperature
Thomas-Fermi framework employing the subtraction procedure. A substantial
decrement in the symmetry energy coefficient is obtained for finite systems,
contrary to those seen for infinite nuclear matter at normal and somewhat
subnormal densities. The effect of the coupling of the surface phonons
to the nucleonic motion is also considered; this is found to decrease 
the symmetry energies somewhat at low temperatures.
\end{abstract}



\maketitle

\section{Introduction}

The nuclear symmetry energy reflects the energy cost in converting the 
isospin asymmetric nuclear matter to the symmetric one. Conventionally,
as part of global mass formulas in the framework of the liquid-drop
model ($\it i.e., $ ignoring pairing and shell-corrections), the
symmetry energy $E_s$ of an even-even or odd-odd finite nucleus
with $N$ neutrons and $Z$ protons  is given from the expression
$E_s = E(N,Z)-E(A/2,A/2)~ =~ e_{sym}\frac {(N-Z)^2}{A} $
where $A=N+Z$, $E(N,Z) $ is the energy of the nucleus barring the Coulomb
part, and $e_{sym} $ is the symmetry energy coefficient. For infinite
nuclear systems, the value of $e_{sym}$ is usually taken in the range
of 30-34 MeV \cite{mye,dan,she}. 
Apart from the evident basic role of symmetry energy
in the correct description of the nuclear binding along the periodic
table and a broad understanding of the nuclear drip lines, it
is of seminal importance in giving a proper description of the dynamical
evolution of the core-collapse of a massive star and the associated explosive
nucleosynthesis. A large (small) magnitude of $e_{sym}$ inhibits (accelerates)
change of protons to neutrons through electron capture \cite{ste,jan}. 
This subtly changes the equation of state (EOS) of the dense stellar matter 
and influences the dynamics of the collapse and 
explosive phase of a massive star.

Stellar matter in the collapse or bounce phase is warm, knowledge about
the temperature dependence of the symmetry energy is therefore a concern
of paramount importance. It is also an important ingredient in properly
accounting for the multiplicity distributions of the fragment masses and
charges and also their isotopic distributions in 
multifragmentation of hot nuclear systems.
Temperature dependence of nuclear symmetry energy and the symmetry
free energy $F_s$ (defined likewise $E_s$ as $F_s~ = ~ F(N,Z)-F(A/2,A/2)
~ = f_{sym}\frac{(N-Z)^2}{A} $, where $F$ is the free energy) for
homogeneous infinite nuclear matter has thence been studied; 
model calculations show 
that $E_s$ for infinite matter 
decreases with temperature while $F_s$ exhibits the 
opposite temperature dependence \cite{xu}. 
The temperature dependence becomes more prominent at low densities, 
particularly for $F_s$.
In stellar core-collapse or bounce 
phase, matter is, however, not homogeneous, it is nucleated \cite{hor,mal} 
and therefore it is important to know how the symmetry energies of finite 
nuclei behave with increasing temperature. The present communication
is aimed to understand this feature.

 Calculations on symmetry energy for finite nuclei in a low temperature
domain ($T \le $ 2 MeV )have been done earlier by Donati $\it et.al $
\cite {don} for applications in core-collapse supernova simulations.
At finite temperature, the symmetry coefficient was found to be
somewhat larger than for cold nuclei ($T$ =0) hindering electron
captures. The calculations were schematic, but takes into account
coupling of the nucleons to the dynamical surface phonons that results
in an increased effective nucleon mass, the so called $\omega$-mass 
$m_{\omega }$ embodying nonlocality in time \cite{has}. This effective mass 
($m_{\omega }$) decreases with temperature, resulting in an increase
in the symmetry energy. Calculations by Dean $\it et.al $ \cite{dea}
in a shell model Monte Carlo (SMMC) framework support these findings.
These calculations have also been done in the same limited temperature
range and may suffer from shortcomings in the isospin dependence 
of the schematic residual interaction they have used. The symmetry
coefficients calculated by them are much below the accepted
nominal values.

  The present calculations have been done in a broader temperature domain
($T \le $ 8 MeV) to have a better understanding of the temperature 
dependence of symmetry energies in atomic nuclei. Finite temperature
Thomas-Fermi framework with the subtraction technique \cite{sur} has been 
employed for this purpose. Two effective interactions, namely the 
modified Seyler-Blanchard (SBM) \cite{ban,de1} and the 
SkM$^*$ \cite{bra} interactions that give 
practically the same symmetry coefficient ($\sim $ 31 MeV) at their respective
saturation densities have been used in our calculations. Dynamical changes
in the energy mass $m_{\omega }$ with temperature is also taken into 
account following the prescription of Refs. \cite{pra,shl}.

The paper is organized as follows. In Sec.II, theoretical framework is
briefly outlined. In Sec. III, results and discussions are given. The
concluding remarks are presented in Sec. IV.

\section{Theoretical framework}

 The methodology employed to calculate the symmetry energy and symmetry
free energy of finite nuclei as a function of temperature is outlined 
in the following.

\subsection{Effective interactions }
In describing the hot nucleus in the finite temperature Thomas-Fermi
(FTTF) approach, we have chosen, as already stated,  
two effective interactions with nearly the same symmetry 
coefficient ($e_{sym} \sim $ 31 MeV at the respective saturation density
for nuclear matter). Both these interactions are  density
dependent and both have a quadratic momentum dependence. The SkM$^*$
interaction has common usage and FTTF calculations with this interaction
\cite{bra,sur,de4} have been performed before, with much success. 
The Seyler-Blanchard \cite{sey} interaction has also been very successfully
used by Myers and Swiatecki \cite{mye1,mye2} in the context of the nuclear
mass formula. The modified version of this interaction (SBM) is found 
to reproduce the ground state binding energies, charge rms radii, 
Giant Monopole Resonance (GMR) energies  etc., 
very well for a host of nuclei from
$^{16}$O to very heavy systems \cite{de1,maj}. The properties of pure
neutron matter calculated with this type of interaction is also seen
to be in good agreement \cite{rud} with those obtained from other
sophisticated interactions.

The SBM  interaction is given by,
\begin{eqnarray}
v_{eff}(r, p,\rho )=C_{l,u} \Bigl [v_1(r,p)
+v_2(r,\rho) \Bigr ], \nonumber \\
v_1=-(1-p^2/b^2)f({\bf r}_1,{\bf r}_2), \nonumber \\
v_2=d^2 \Bigl [ \rho(r_1)+\rho (r_2) \Bigr ]^\kappa 
f({\bf r}_1,{\bf r}_2),
\end{eqnarray}
with
\begin{eqnarray}
f({\bf r}_1,{\bf r}_2)=\frac {e^{-|{\bf r}_1-{\bf r}_2|/a}}
{|{\bf r}_1-{\bf r}_2|/a}.
\end{eqnarray}
The isospin dependence in the interaction is brought through the different
strength parameters $C_l$ for like-pairs (n-n,p-p) and $C_u$ for unlike
pairs (n-p). The relative separation of the nucleons in configuration
and momentum space are given by $r=|{\bf r_1-r_2 }|$ and $p=|{\bf p_1-p_2 }|$.
The densities at the sites of the two interacting nucleons are given by 
$\rho ({\bf r_1})$ and $\rho ({\bf r_2})$. The range of the interaction
is $a, b$ and $d$ are the measures of the momentum and density dependence
in the interaction and $\kappa $ controls the stiffness of the nuclear 
EOS.

The five parameters $C_l, C_u, a, b,$ and $d$ for a fixed value of the
density exponent $\kappa $ are determined by reproducing  (i) the volume
energy coefficient for symmetric nuclear matter ($a_v = -$16.1 MeV),
(ii) saturation density of normal nuclear matter (taken as $\rho_0 =$
0.1533 fm$^{-3}$, corresponding to the radius parameter $r_0 =$ 1.16 fm),
(iii) the volume symmetry coefficient $e_{sym} $= 31 MeV,
(iv) the surface energy coefficient of symmetric nuclear matter 
($a_s$ =18.01 MeV), and (v) the energy dependence of the real part $V_{An}$
of the nucleon-nucleus optical potential ($\frac {dV_{An}}{dE} =$ 0.30).

The procedure for determining the parameters are given in detail in Ref.
\cite{ban}. The parameter $\kappa $ is obtained by reproducing the giant
monopole resonance energies of a large number of nuclei employing the 
scaling model \cite{maj}. The parameters of the interaction are listed 
in Table. I. These parameters differ somewhat from the ones used
in Ref. \cite{de1} where the volume symmetry coefficient 
$e_{sym} $was taken as 34 MeV. In the present communication,
it is taken as 31 MeV,  a value commonly
used by several authors  with some justification from experimental
findings \cite {she}. The effective $k$-mass of the nucleon $m_k$
coming from the momentum dependence of the effective interaction,
comes out to be 0.62$m$ for symmetric matter, where $m$
is the nucleon mass. The isoscalar volume incompressibility $K_{\infty }$,
symmetry incompressibility $K_{sym} 
[ = (9 \rho_0^2 \frac {\partial ^2 
e_{sym}}{\partial \rho ^2})_{\rho_0} ] $ and the symmetry pressure
$L [=(3 \rho_0 \frac {\partial e_{sym}}{\partial \rho })_{\rho_0} 
] $ are
240, $-$101 and 59.8 MeV, respectively for the SBM interaction.
The corresponding values for SkM$^*$ interaction are 216.7, $-$155.9 and
45.8 MeV.

\begin{table}
\caption{ The parameters of the SBM effective interaction (in MeV fm units)}
\begin{ruledtabular}
\begin{tabular}{cccccc}
$C_l$&  $C_u$& $a$& $b$& $d$& $\kappa $\\
\hline
348.5& 829.7& 0.6251& 927.5& 0.879& 1/6\\
\end{tabular}
\end{ruledtabular}
\end{table}

\subsection{Describing the hot nucleus}

In the FTTF approach, the nucleon density profile for the hot
nucleus is arrived at self-consistently. The details for obtaining
the density in this method for the SkM$^*$ 
force are given in Ref. \cite{bra};
for the SBM interaction, they have also been presented earlier
\cite{de1,sam1}. However, for completeness, we present the salient
features below.

The total energy of the nucleus is given by 
\begin{eqnarray}
E&=&\int d{\bf r}[\varepsilon_K(r)+\varepsilon_I(r)+\varepsilon_c(r)]
\nonumber \\
&&=E_K+E_I+E_c
\end{eqnarray}
where $\varepsilon_K(r), \varepsilon_I(r)$ and $\varepsilon_c(r)$
are the kinetic, nuclear interaction and Coulomb interaction energy densities,
respectively; $E_K$, $E_I$ and $E_c$ are the corresponding total energies. 
The energy densities are given by
\begin{eqnarray}
\varepsilon_K(r)=\frac {2}{h^3}\sum_\tau \int d{\bf p}\frac{p^2}{2m_\tau}
n_\tau (r,p,T)
\end{eqnarray}
\begin{eqnarray}
\varepsilon_I(r)&=&\frac{2}{h^3}\sum_\tau \Bigg [\frac{1}{h^3}
\int \Big \{v_1(|{\bf r}-{\bf r^\prime }|,|{\bf p}-{\bf p^\prime }|) 
\nonumber \\
&&+v_2(|{\bf r}-{\bf r^\prime }|,\rho )\Big \} 
 \Big \{ C_ln_\tau (r^\prime ,p^\prime)+C_u
n_{-\tau }(r^\prime ,p^\prime )\Big \} \nonumber \\
&&\times n_\tau ( r,p)
\Bigg ]d{\bf r}^\prime d{\bf p}d{\bf p}^\prime
\end{eqnarray}
\begin{eqnarray}
\varepsilon_c(r)&=&e^2\pi \rho_p(r)\int dr^\prime r^{\prime^2}
\rho_p(r^\prime)g(r,r^\prime ) \nonumber \\
&& -\frac{3e^2}{4\pi }(3\pi^2)^{1/3}\rho_p^{4/3}(r)~.
\end{eqnarray}
In these equations, $n_\tau (r,p,T) $ is the nucleon distribution function.
In the Coulomb energy density, the first and second terms are the direct
and exchange parts, respectively. Here, $\rho_p(r)$ is the proton density
profile and 
\begin{eqnarray}
g(r,r^\prime )=\frac {(r+r^\prime )-|r-r^\prime |}{rr^\prime }.
\end{eqnarray}
In Eqs.~(4) and (5), $\tau$ is the isospin index; if $\tau $ refers to
a proton, $-\tau$ refers to a neutron and {\it vice versa}; 
the distribution function $n_\tau $ is  
determined from the minimization of the
total thermodynamic potential $\Omega$ ($=E-TS-\sum_\tau \mu_\tau 
N_\tau $) \cite{de1}. Here $S$ is the total entropy of the system
that can be obtained from Landau quasiparticle approximation as,

\begin{eqnarray}
S=-\frac {2}{h^3} \sum_\tau \int  \Bigl [ n_\tau \ln n_\tau +
(1-n_\tau )\ln (1-n_\tau ) \Bigr ] d{\bf r}d{\bf p},
\end{eqnarray}
and $\mu_\tau $ are the nucleon chemical potentials. The distribution
function is given as

\begin{eqnarray}
n_\tau (r,p,T)= exp \Bigl [ \Bigl (\frac {p^2}{2m_\tau }+V_\tau ^0 +
p^2 V_\tau ^1 +V_\tau ^2 -\mu_\tau \Bigr ) \Bigr ]^{-1}
\end{eqnarray}
In Eq.~(9), $V_\tau ^0(r)+p^2V_\tau ^1(r) $ 
is the single-particle potential
and $V_\tau ^2(r) $ is the rearrangement potential originating from the
density dependence of the interaction. The full expressions for the 
different components of the single-particle potential and $V_\tau ^2$
are given in Ref. \cite{de1} and are not repeated here. The effective
$k$-mass of the nucleon $m_{\tau ,k}(r)$ can be defined as
\begin{eqnarray}
m_{\tau ,k}=\Bigl [ \frac {1}{m_\tau }+2V_\tau ^1 \Bigr ]^{-1}.
\end{eqnarray}
From Eq.~(9), the nucleon density
\begin{eqnarray}
\rho_\tau (r)=\frac {2}{h^3}\int n_\tau(r,{\bf p}, T)d{\bf p} 
\end{eqnarray}
is obtained as,
\begin{eqnarray}
 \rho_\tau (r)= A_T^* (r) J_{1/2}[\eta_\tau (r)],
\end{eqnarray}
where 
\begin{eqnarray}
 A_T^*(r)=\frac{4\pi}{h^3}[2m_{\tau ,k}(r)T]^{3/2},
\end{eqnarray}
and $J_k(\eta )$ is the Fermi integral 
\begin{eqnarray}
J_k(\eta )=\int_0^\infty \frac{x^k}{1+exp(x-\eta )} dx
\end{eqnarray}
with the fugacity  $\eta $ given as
\begin{eqnarray}
\eta _\tau (r) = [\mu_\tau -{\cal V}_\tau (r) ]/T.
\end{eqnarray}
Here  ${\cal 
V}_\tau (r)$ is the effective single-particle (SP) potential (Coulomb
included), defined as
\begin{eqnarray}
{\cal V}_\tau (r)=\Bigl ( V_\tau ^0(r)+V_\tau ^2 (r) +{\delta }_{\tau ,p}
V_c(r) \Bigr )/T,
\end{eqnarray}
$V_c(r)$ being the single-particle Coulomb potential,
\begin{eqnarray}
V_c(r)&=&2e^2\pi \int dr^\prime {r^\prime }^2 \rho_p (r^\prime )
g(r,r^\prime ) \nonumber \\
&& -\frac {e^2}{\pi}(3\pi^2)^{1/3} \rho_p(r)^{1/3}.
\end{eqnarray}

From Eqs.~(4) and (5), the kinetic and nuclear interaction energy
densities can be simplified to
\begin{eqnarray}
{\varepsilon }_K (r)= \frac{2\pi}{h^3}\sum_\tau \frac{1}{m_\tau }
(2m_{\tau ,k}T)^{5/2}J_{3/2}(\eta_{\tau } (r))
\end{eqnarray}
\begin{eqnarray}
{\varepsilon }_I (r)&=& \frac {1}{2} 
\sum_\tau \rho_\tau (r) V_\tau ^0 (r) \nonumber \\
&&+\frac {2\pi }{h^3}\sum_\tau (2 m_{\tau ,k}T)^{5/2}V_\tau ^1(r) 
{J_{3/2} (\eta_\tau (r))}.
\end{eqnarray}
For a finite nucleus, the total energy density is the sum of 
Eqs.~(6), (18) and (19). For infinite
systems, the energy densities are space independent. For a
particular density $\rho $, the kinetic and potential energies
per nucleon can then be defined as $e_K=\varepsilon_K/\rho $
and $e_I=\varepsilon_I/\rho $, respectively.

At a large distance from the center, particularly for neutrons, 
${{\cal V}_\tau } \sim $ is zero and the
density becomes $\rho_{\tau} \sim e^{\mu_{\tau} /T} $ 
which is zero at $T =$ 0 as
$\mu_{\tau} $ is negative. At  finite $T$, however, 
$\rho_{\tau} $ is not zero and
takes a constant value with a nonzero pressure at the surface. This
makes the system thermodynamically unstable. Moreover, the density
profile becomes dependent on the size of the box in which calculations 
are done. This problem is overcome in the subtraction procedure \cite{sur,bon},
where the hot nucleus, assumed to be a thermalized system in equilibrium
with a surrounding gas simulating the effects of
the evaporated nucleons, is separated
from the embedding environment. This method is based on the existence
of two solutions to the FTTF equations, one corresponding to the liquid
phase with the surrounding gas ($lg$) and the other corresponding to
the gas ($g$) phase. The density profile of the hot nucleus in 
thermodynamic equilibrium is given by $\rho_{\tau ,l}(r)=\rho_{\tau ,lg} (r)  
-\rho_{\tau ,g}(r)$ and it is independent of the choice of the box size.
The conservation of nucleon number of each species $N_\tau $ of the
hot nucleus gives
\begin{eqnarray}
\int [\rho_{\tau ,lg}(r)-\rho_{\tau ,g} (r)]d{\bf r } = N_\tau.
\end{eqnarray}

The energy $E$ of the nucleus is given by 
\begin{eqnarray}
E=E_{lg}-E_{g},
\end{eqnarray}
where $E_{lg}$ and $E_g$ are the total energies of the liquid-gas
system and the gas alone. From Eq.~(8), the total entropy 
could be recast as \cite{de2},
\begin{eqnarray}
S=-\sum_\tau \int g_\tau (\varepsilon _\tau ,T)[f_\tau \ln f_\tau +
(1-f_\tau )\ln (1-f_\tau )]d \varepsilon _\tau ,
\end{eqnarray}
where $f_\tau $ is the single-particle occupation function
\begin{eqnarray}
f_\tau (\varepsilon _\tau ,\mu_\tau ,T)~=~[1+exp{(\varepsilon _\tau
-\mu_\tau )/T}]^{-1},
\end{eqnarray}
and $g_\tau $ is the subtracted single-particle level density. It is
given as \cite{sam1},
\begin{eqnarray}
g_\tau (\varepsilon _\tau ,T)&=&\frac{4\sqrt 2}{\pi \hbar ^3}
\int \Bigl [(m_{\tau ,k}^{lg})^{3/2} \sqrt {\varepsilon _\tau -{\cal V}_\tau
^{lg}(r)} \nonumber  \\
&& - (m_{\tau ,k}^g)^{3/2} \sqrt {\varepsilon _\tau -{\cal V}_\tau^g
(r)} \Bigr ]r^2 dr,
\end{eqnarray}
where ${\cal V}_\tau ^i $ is the single-particle potential, $i$ referring
to $lg$ or $g$.

The free energy is calculated from $F=E-TS$. In terms of the occupation
function, the density in the FTTF approximation could be written as 
\begin{eqnarray}
\rho_\tau ^i(r)&=&\frac{1}{2\pi ^2 \hbar ^3}(2m_{\tau ,k}^i)^{3/2}
\int \sqrt {\varepsilon _\tau-{\cal V}_\tau ^i } \nonumber  \\
&&~ \times f(\varepsilon _\tau ,
\mu_\tau ,T) d\varepsilon _\tau .
\end{eqnarray}

\subsection{The energy-mass }

The effective mass $m^*$ has two components, the $k$-mass and the energy
mass (or the $\omega $-mass); it is defined as
\begin{eqnarray}
m^*~=~m(\frac{m_k}{m})(\frac{m_\omega }{m}),
\end{eqnarray}
where $m$ is the nucleon mass.

The temperature-dependent $k$-mass is given by the momentum-dependent
part of the single-particle potential as discussed earlier. 
The $\omega $-mass $m_\omega $
originates from  the coupling of the
single-particle motion with the collective degrees of freedom. The
$m_\omega $ varies with position and temperature.
A self-consistent calculation of the temperature-dependent $m_\omega $
is very involved and not within the scope of the present work. We have 
therefore taken a phenomenological form \cite{shl} for $m_\omega $
such that
\begin{eqnarray}
\frac{m_\omega }{m}~=~1-0.4A^{1/3}exp \Bigl [
- \Bigl (\frac {T}{21A^{-1/3}} \Bigr )^2 \Bigr ]
\frac{1}{\rho (0)}\frac {d \rho (r)}{dr}.
\end{eqnarray}
The temperature $T$ and distance $r$ are measured in MeV-fm units,
$\rho (0)$ is the central density of the nucleon distribution in the
nucleus. The collectivity refers to the liquid phase only. The density in
the above equation is $\rho (r)~=~\rho_{lg} (r)-\rho_g (r) $;  
$A$ refers to the liquid mass. This implies $m_\omega ^{lg}~=~m_\omega $
and $m_\omega ^g~=~m$.

To avoid the complexities arising from the self-consistent calculation of the
density profile with the inclusion of $\omega $-mass, we have adopted a 
realistic extension of the method given in Ref. \cite {shl}. This is
described in some detail in Ref. \cite {de2}.

The subtracted level density corresponding to Eq.(24) then modifies to
\begin{eqnarray}
\tilde g_\tau (\varepsilon _\tau ,T)&=&\frac{4\sqrt 2}{\pi \hbar^3}
\int  \Bigl [(m_{\tau ,k}^{lg} \frac {m_\omega }{m})^{3/2} 
\sqrt {\varepsilon _\tau -{\cal V}_\tau ^{lg}(r) 
\frac{m}{m_\omega }} \nonumber \\
&&-(m_{\tau ,k}^g)^{3/2} \sqrt {\varepsilon _\tau -{\cal V}_\tau ^g (r)} \Bigr ]
r^2dr ,
\end{eqnarray}
the densities in the $lg$ or $g$ phase modify accordingly,
\begin{eqnarray}
\tilde \rho_\tau ^i(r)&=&\frac{1}{2\pi ^2 \hbar ^3} \Bigl [2m_{\tau ,k}^i
\frac{m_\omega ^i}{m} \Bigr ]^{3/2}
 \int \sqrt {\varepsilon _\tau
-{\cal V}_\tau ^i \frac{m}{m_\omega ^i}} \nonumber  \\
&& \times ~f(\varepsilon _\tau ,\tilde \mu_\tau ,T)
d\varepsilon _\tau .
\end{eqnarray}
The chemical potential $\mu_\tau $ has modified to $\tilde \mu_\tau $
to conserve the particle numbers in the nucleus
\begin{eqnarray}
N_\tau~=~\int \tilde g_\tau (\varepsilon _\tau ,T)f_\tau (\varepsilon _\tau ,
\tilde \mu_\tau ,T) d \varepsilon _\tau .
\end{eqnarray}

\subsection{The symmetry coefficients}

For nuclear matter, the symmetry coefficient at density $\rho $
can be defined as,
\begin{eqnarray}
e_{sym}(T)= [ e(\rho ,X,T)-e(\rho ,X=0,T) ]/X^2,
\end{eqnarray}
where $e$ is the energy per nucleon  and $X$ is the asymmetry
parameter $X=(\rho_n-\rho_p)/\rho )$. Here $\rho_n$ and $\rho_p$
are the neutron and proton densities $(\rho =\rho_n+\rho_p)$.The
kinetic and potential parts of the symmetry coefficients 
$e_{sym}(k)$ and $e_{sym}(v)$  can likewise be defined replacing
$e$  in the square brackets in Eq.~(31) by $e_K$ and $e_I$, respectively.

For a nucleus of mass $A$, the symmetry coefficient, in the spirit
of the liquid-drop model, can also be defined as
\begin{eqnarray}
e_{sym}(T)=[e(N,Z,T)- e(A/2,A/2,T)]/X^2.
\end{eqnarray}
Here the asymmetry parameter is $X=(N-Z)/A$. In this equation, $e$ is the
energy per particle of the nucleus barring the 
Coulomb contribution. For relatively
heavy nuclei, stable systems are usually isospin-asymmetric, there the
definition given by Eq.(32) may not be operative. In such systems, $e_{sym}$
could be calculated from 
\begin{eqnarray}
e_{sym}(T)=[e(A,X_1,T)-e(A,X_2,T)]/(X_1^2-X_2^2),
\end{eqnarray}
where $X_1$ and $X_2$ are the asymmetry parameters of the nuclear
pair. Similar definitions follow for symmetry free energy coefficient
$f_{sym}$ on replacing energy $e$ by the free energy per particle $f$.

 The value of the symmetry coefficient obtained from Eqs.~(32) and
(33) depends on the choice of the nuclear pair, its value is therefore
not unambiguous for a particular nucleus. However, the local density
approximation (LDA) could be used to define  the symmetry
coefficient for a specific nucleus once its density profile is
known. In the LDA, it is given as \cite{sam2}
\begin{eqnarray}
e_{sym}(T) \Bigl (\frac{N-Z}{A} \Bigr )^2&=&\frac{1}{A}
\int \rho (r)e_{sym}^\infty 
[\rho (r),T] \nonumber \\ 
&&\times ~ \Bigl [\frac {\rho_n (r)-\rho_p (r)}{\rho (r)} \Bigr ]^2 d{\bf r}.
\end{eqnarray}
Here $e_{sym}^\infty [\rho (r),T]$ is the symmetry energy coefficient
of infinite matter at temperature $T$ at a value of the density $\rho (r)$. 
The local isospin density is given by $[\rho_n (r)-\rho_p (r)]$.

\section{Results and discussions}

As already stated, for the microscopic finite temperature Thomas-Fermi
calculations, we have chosen the SkM$^*$ and the SBM effective interactions.
Both these interactions reproduce the ground state properties of atomic
nuclei and also of nuclear matter fairly well. For normal nuclear matter
at $T$=0,
these interactions have symmetry coefficient $e_{sym} \sim $ 31.0 MeV,
a commonly accepted value  with some support from experimental
analyses \cite{she}. Below the saturation density, both these interactions
display nearly the same density dependence for $e_{sym}$ 
as shown in Fig.~1. The
symmetry coefficient obtained from the SBM interaction 
(given by the red line) looks a 
little stiffer beyond the saturation density. It would be interesting
to compare the symmetry coefficients of cold nuclei obtained from
these interactions with phenomenological values. This is shown in
panels (a) and (b) of
Fig.~2; the full green lines in both the panels
correspond to the upper and lower bounds for
$e_{sym}$ as given in Ref.~\cite{dan} as
\begin{eqnarray} 
e_{sym}=\frac {\alpha }{1+\frac {\alpha }{\beta }A^{-1/3}},
\end{eqnarray} 
where $\alpha $= 31.0 MeV and $\frac {\alpha }{\beta } $ =2.4 $\pm $0.4.
The black circles and red triangles refer to calculations with SkM$^*$
and SBM interactions, respectively. 
The upper panel corresponds to calculations done in the difference
method using Eq.~(33). The nuclear pairs $(A,Z_1)$ and $(A,Z_2)$
collectively written as $(A,Z_1,Z_2)$ are chosen as (26,10,12),
(40,16,18), (56,24,26), (80,34,36), (120,50,52),(150,60,62),
(197,77,79),and (238,90,92). These results are representative,
the values of $e_{sym}$ depend on the choice of nuclear pairs as shown later.
The lower panel displays results calculated in the LDA using Eq.~(34).
The nuclei $(A,Z)$ are chosen close to the $\beta$-stability line; 
they are (26,12), (40,18), (56,26), (80,34), (120,50), (150,62),
(197,79), and (238,92), respectively. In the LDA, the values of
$e_{sym}$ are seen to have a relatively weak dependence on the
choice of the asymmetry parameter $X$.
Compared to those in SBM, the
SkM$^*$ values are somewhat lower. 
The values of $e_{sym}$, however, are seen to lie within the phenomenological
limits, particularly in the LDA.
The difference in results in $e_{sym}$
for finite nuclei in spite of the apparent similarity in its density
dependence for infinite nuclear matter stems from the fact  that
the SBM is a finite range interaction as opposed to SkM$^*$ 
(zero range) which results in  somewhat different density profiles
of nuclei.

\begin{figure}
\includegraphics[width=1.0\columnwidth,angle=0,clip=true]{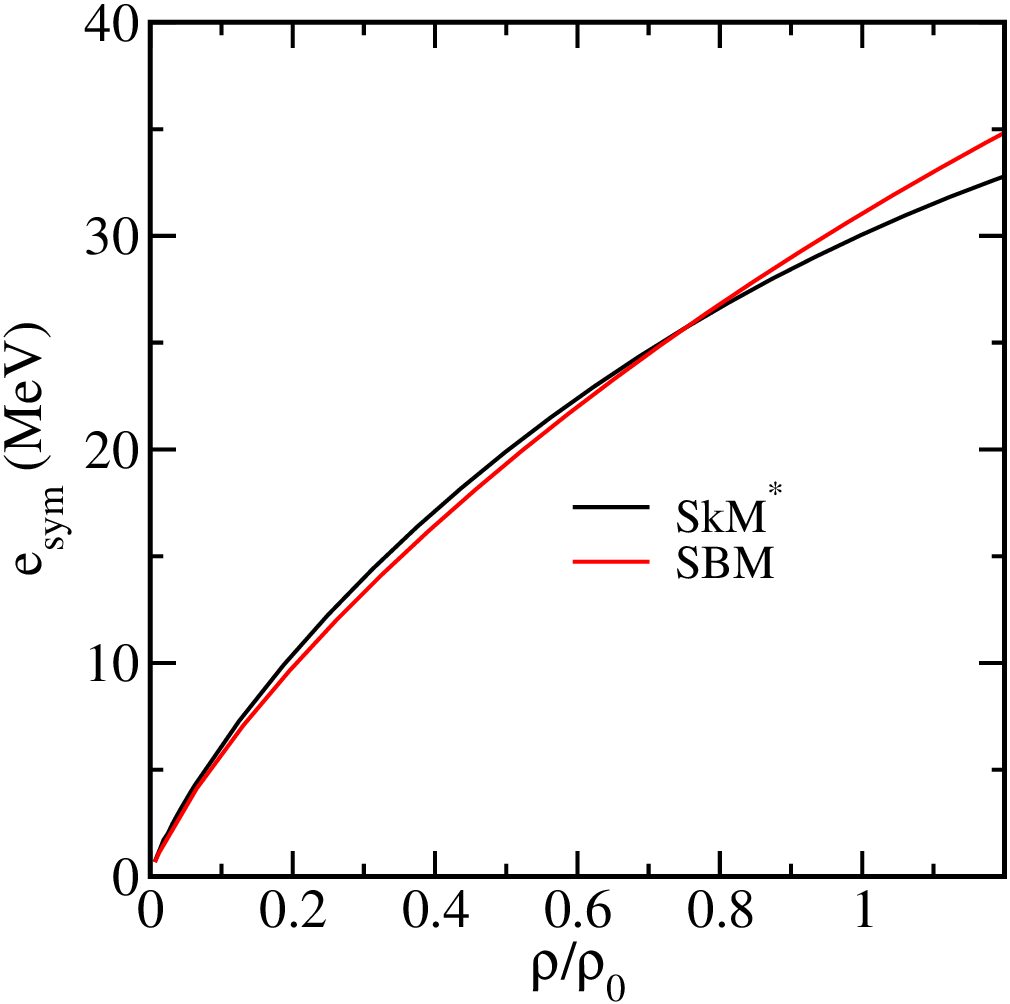}
\caption{(color online)The symmetry energy coefficient $e_{sym}$
for nuclear matter at $T=$0 as a function of density with SkM$^*$ (black) and
SBM (red) interactions.} 
\end{figure}

\begin{figure}
\includegraphics[width=1.0\columnwidth,angle=0,clip=true]{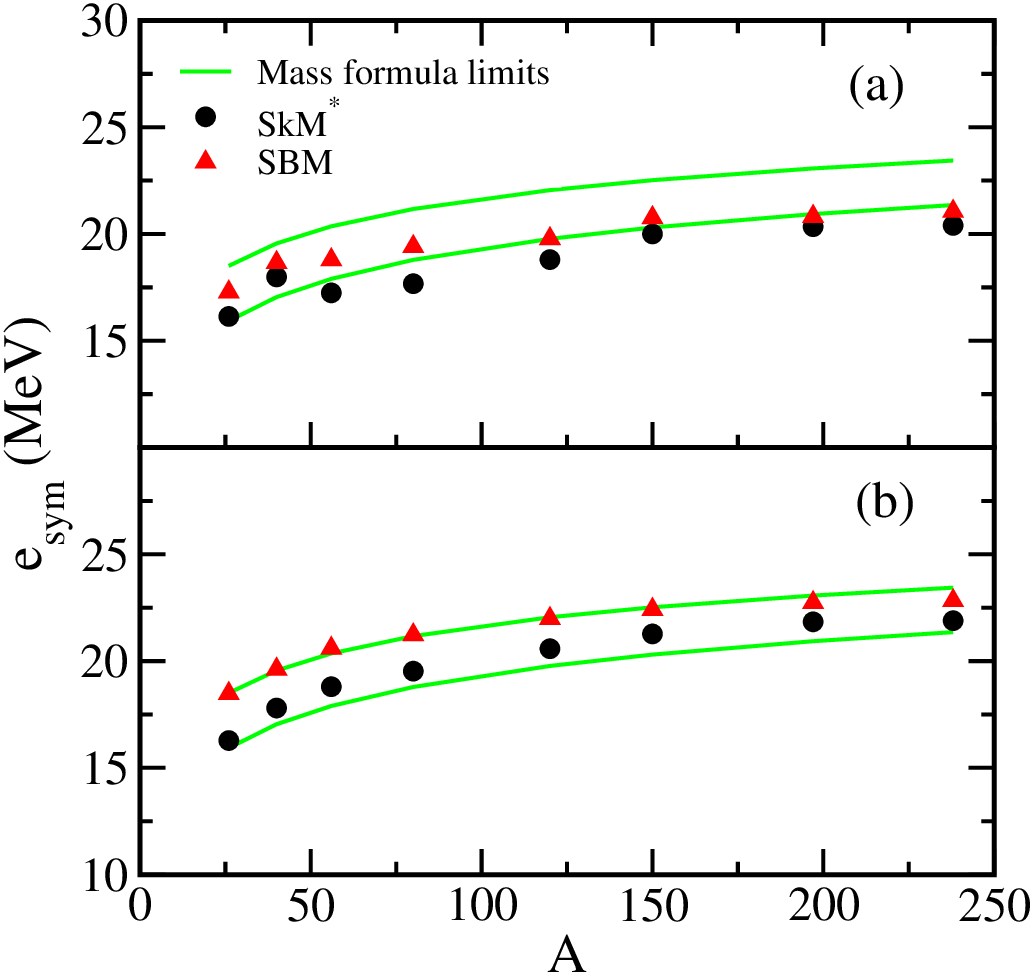}
\caption{(color online)The dependence of symmetry energy coefficient 
$e_{sym}$ on nuclear mass number with SkM$^*$ (black circles) and
SBM (red triangles) interactions.  The limits in the nuclear mass formula
as given in Ref. \cite{dan} are also shown (green lines). The upper and
lower panels correspond to calculations using Eqs.~(33) and (34),
respectively. For details, see text.} 
\end{figure}

Before embarking on the evaluation of the temperature dependence of the
symmetry coefficients for finite systems, it may be worthwhile to
see how $e_{sym}$ or $f_{sym}$ behave with increasing temperature for
infinite nuclear matter. At saturation density, we find that both
$e_{sym}$ and $f_{sym}$, in the temperature range we have studied, 
have a very weak temperature dependence with both the interactions
chosen. At relatively low densities, $e_{sym}$ is seen to decrease with
temperature, in consonance with that reported earlier by Xu
et.al., \cite{xu} and Moustakidis \cite{mou}. At these densities,
$f_{sym}$, however, displays a comparably more prominent rise,
in fair agreement with those obtained earlier \cite{sam2,xu}.
 This is displayed
in Fig.~3  at $\rho =\rho_0/8$ 
($\rho_0=$ saturation density) for the SBM
interaction. Results with SkM$^*$ interaction do not show any different
behavior and are therefore not shown. The  fall in $e_{sym}$ 
(shown by the red line) with
temperature at the low density is essentially due to the decrement
in its kinetic energy part $e_{sym}(k)$ (shown by the green line). 
This part of the symmetry
energy decreases at lower densities as Pauli-blocking becomes less important
because of increased diffuseness of nucleon Fermi surfaces with
temperature. The potential component  $e_{sym}(v)$ 
(black line) decreases more slowly.  
The increase of $f_{sym}$ with temperature can be understood from the
fact that
\begin{eqnarray} 
f_{sym}~=~ e_{sym}-T\frac{s_{sym}}{X^2}~,
\end{eqnarray} 
where $s_{sym}$, the symmetry entropy is negative and is $\sim -\frac
{1}{2}X^2$ \cite{de3} for dilute nuclear matter. Thus even if
$e_{sym}$ falls slower, the increase in $f_{sym}$ (lower panel) 
is  noticeable.

\begin{figure}
\includegraphics[width=1.0\columnwidth,angle=0,clip=true]{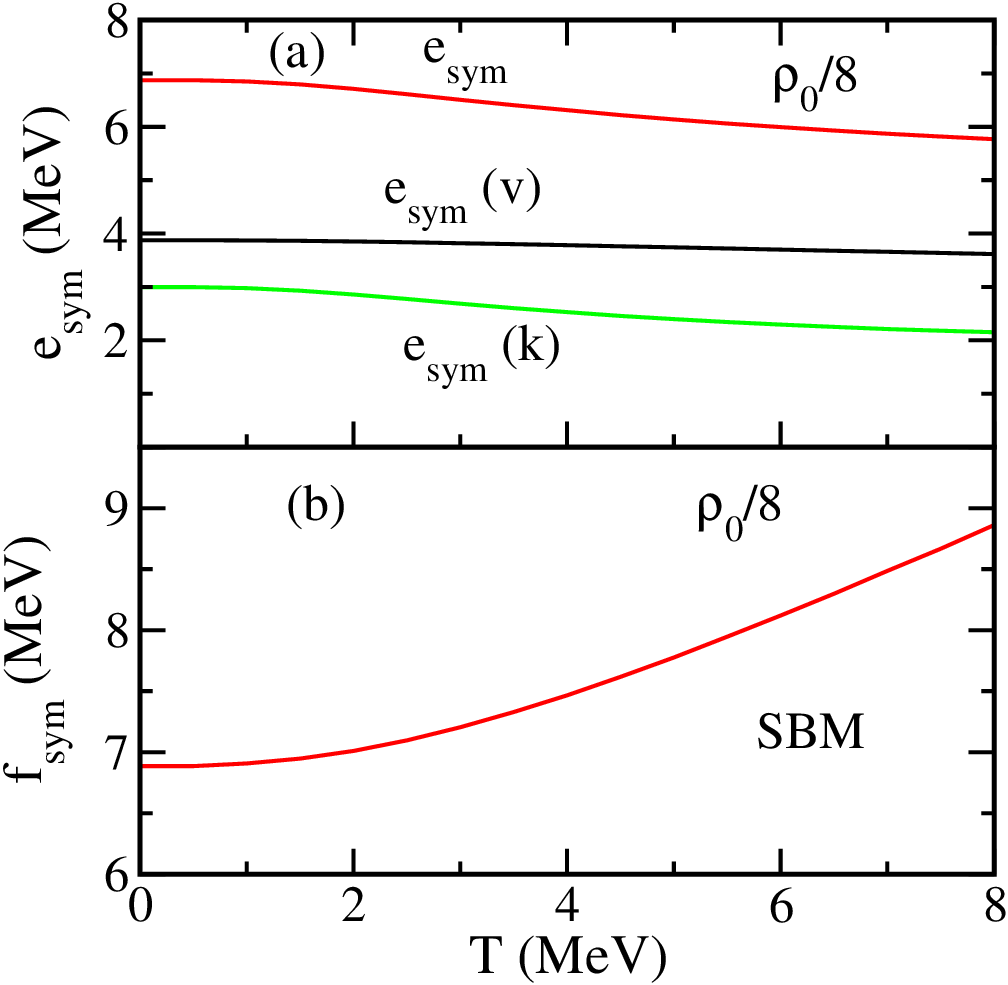}
\caption{(color online) In the upper panel
the temperature evolution of kinetic (green) and 
potential (black) components $e_{sym}(k)$ and $e_{sym}(v)$, respectively,  
of symmetry energy coefficient  along with their sum $e_{sym}$ (red)
are shown for nuclear matter at a density $\rho_0 /8$ calculated with SBM 
interaction. In the lower panel, the same is shown for the total
symmetry free energy coefficient (red) $f_{sym}$.}
\end{figure}
 
\begin{figure}
\includegraphics[width=1.0\columnwidth,angle=0,clip=true]{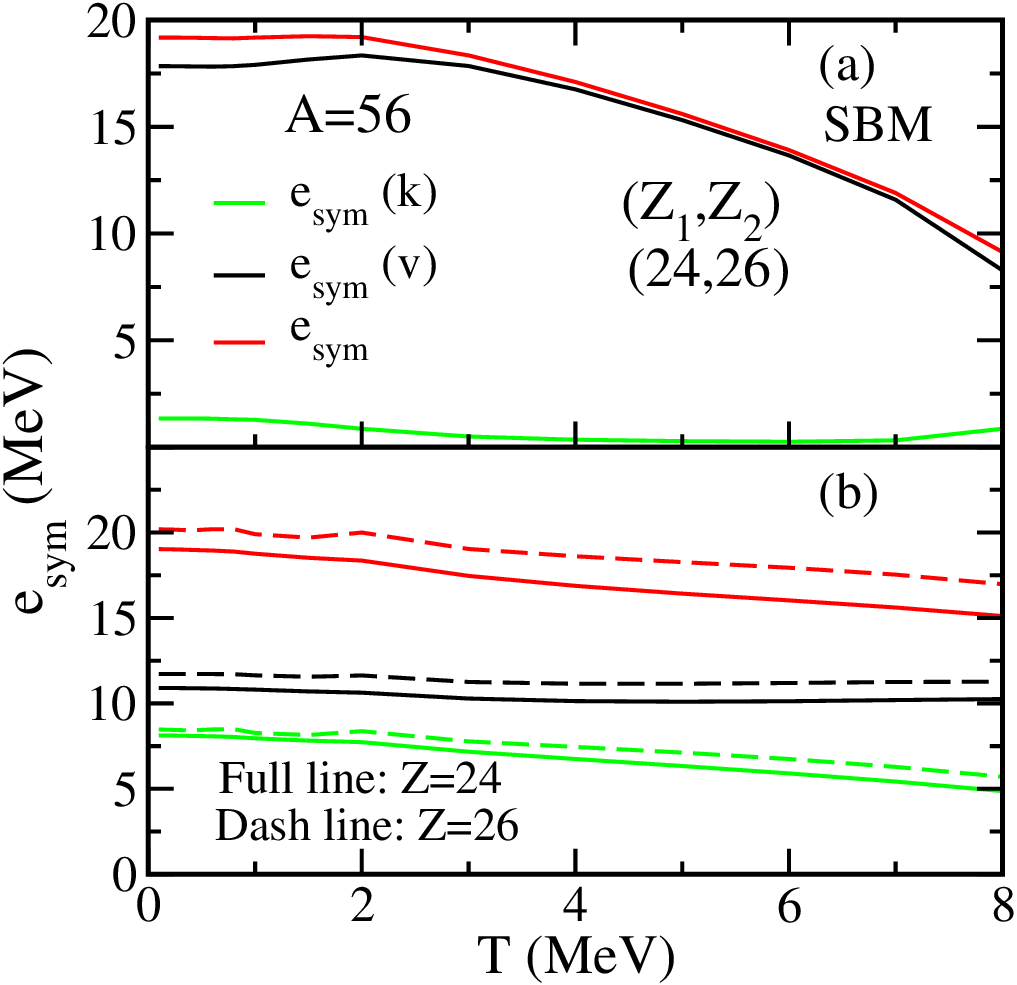}
\caption{(color online) The upper panel represents
the temperature evolution of kinetic (green) and 
potential (black) components   
of symmetry energy coefficient  along with their sum $e_{sym}$ (red)
obtained from A=56 isobaric pair with Z=24 and 26 using SBM 
interaction. In the lower panel, the same is shown for these isobaric
nuclei individually in the local density approximation.}
\end{figure}

\begin{figure}
\includegraphics[width=1.0\columnwidth,angle=0,clip=true]{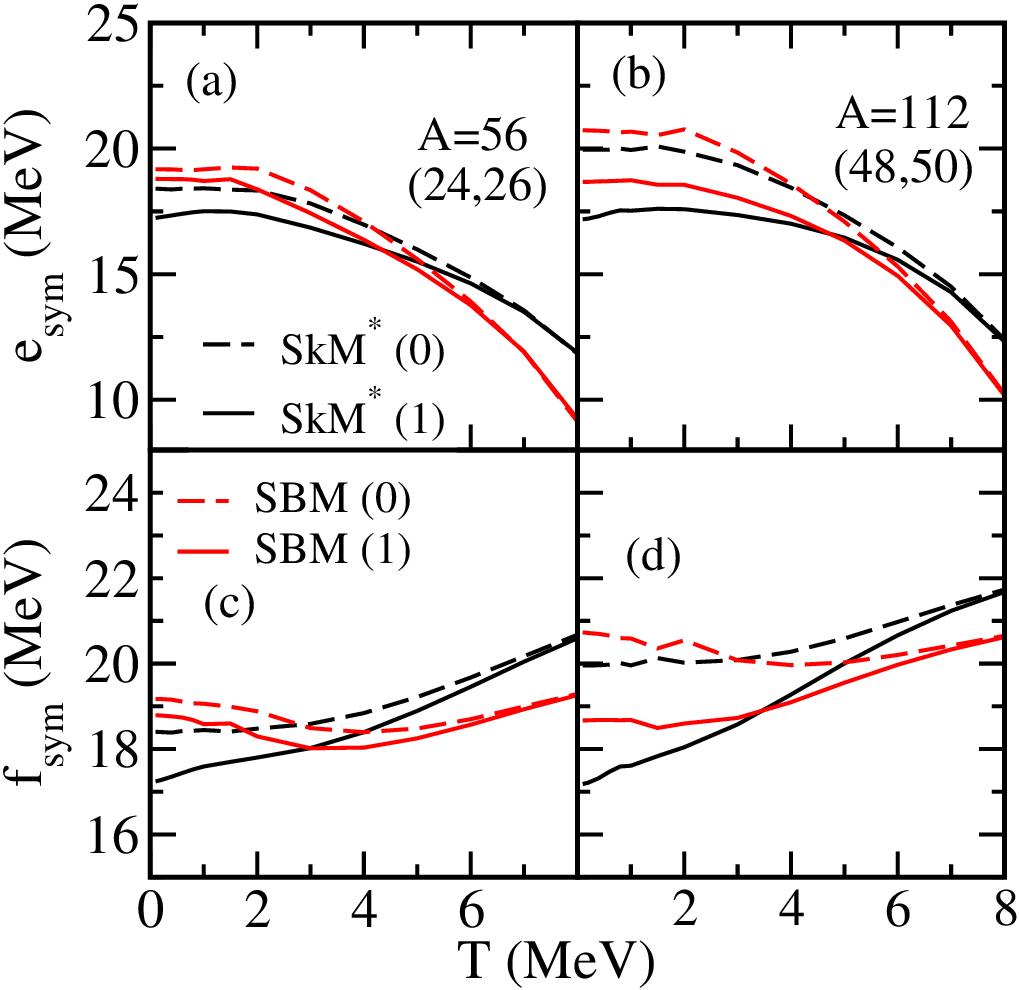}
\caption{(color online) The temperature dependence of $e_{sym}$
using SkM$^*$ interaction with (SkM$^*$(1)) and without (SkM$^*$(0))
inclusion of $\omega$-mass are shown in panel (a) by black full and
dash lines, respectively, for A=56 calculated with Z=(24,26) isobaric pair.
The red lines represent the same for SBM interaction. The same is shown
in panel (b) for A=112 using Z=(48,50) isobaric pair. The panels (c) and
(d) represent the same for $f_{sym}$.}
\end{figure}

\begin{figure}
\includegraphics[width=1.0\columnwidth,angle=0,clip=true]{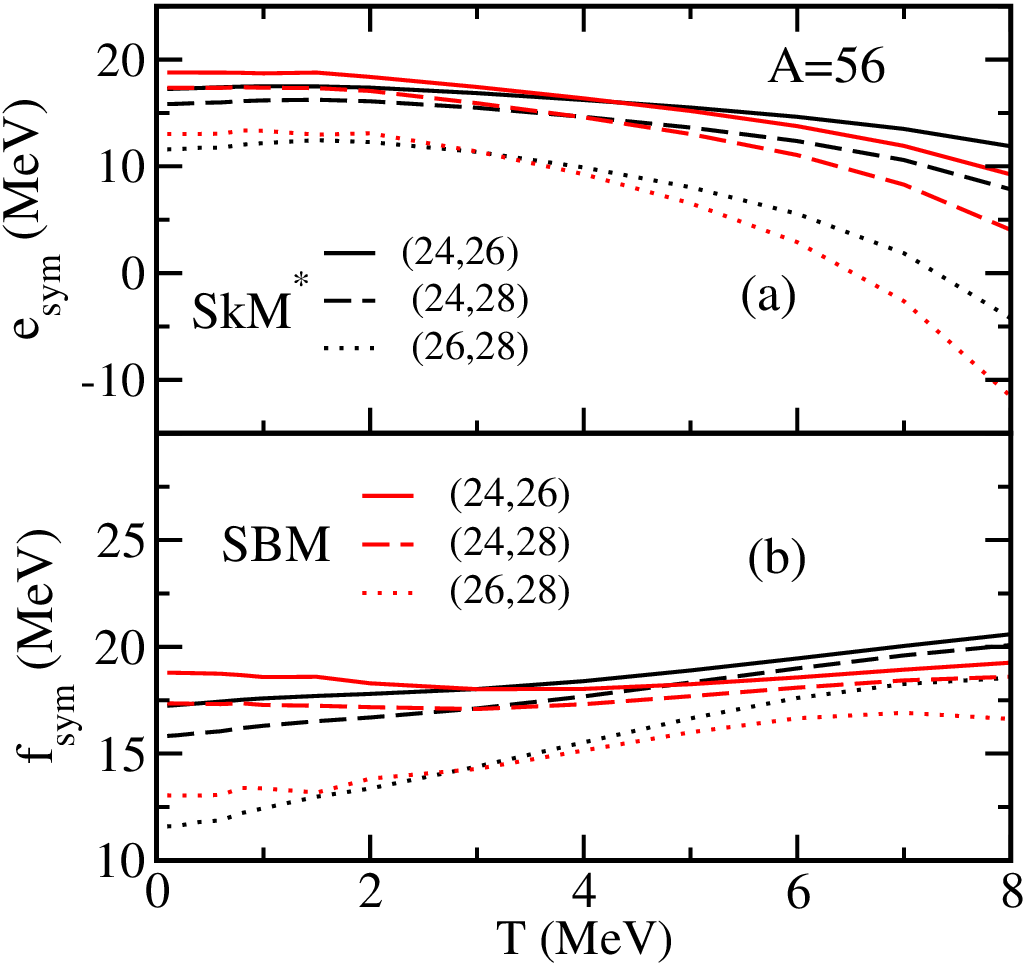}
\caption{(color online) The temperature dependence of $e_{sym}$ 
(upper panel) and $f_{sym}$ (lower panel) for the nucleus A=56.  
The black and red lines correspond to results with SkM$^*$ and
SBM interaction, respectively. The numbers in the parentheses
refer to charge numbers of the different isobaric pairs
used to calculate the symmetry energies.}  
\end{figure}

In Fig.~4, the temperature dependence of the symmetry energy coefficients
and their kinetic and potential components for a nucleus $A$=56 is shown
for the SBM interaction. The upper panel represents calculations in the 
difference method where Eq.~(33) is exploited to obtain the symmetry
coefficients with $Z_1=$24 and $Z_2=$26. 
The density profiles of the nuclei have been calculated with Coulomb
switched on, however, the symmetry coefficients are calculated from
the energies obtained after subtraction of the Coulomb contribution.
The lower panel shows the 
calculations in LDA with the use  of Eq.~(34) for the nuclei with
$Z=$24 and 26. The difference method shows that for nuclei, the main
contribution to $e_{sym}$ (shown by the red line)
comes from the potential part (the black line), $e_{sym}(k)$ (the green line)
is quite small. The symmetry coefficient is nearly constant upto $\sim $
2 MeV, but then falls substantially with temperature. The calculations in 
the LDA, for the isobars with $Z=$ 24 and 26, however, show a very weak
temperature dependence for the symmetry coefficients, which comes mostly
from the surface region; the temperature dependence of infinite matter
is reflected in these calculations for finite nuclei. Here 
the contributions from $e_{sym}(v)$ and $e_{sym}(k)$ are comparable.

\begin{figure}
\includegraphics[width=1.0\columnwidth,angle=0,clip=true]{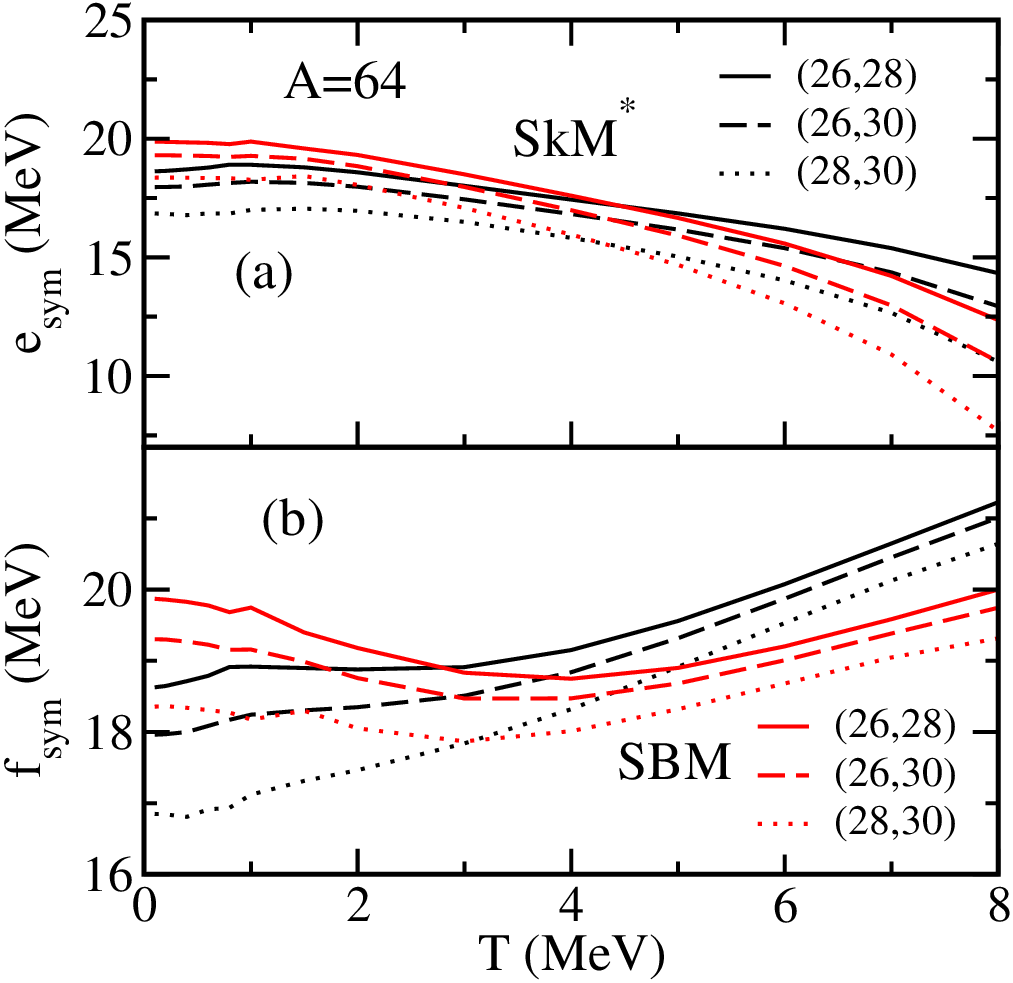}
\caption{(color online) The same as in Fig.~6 for the nucleus A=64
with isobar pairs as shown in the parentheses.} 
\end{figure}

\begin{figure}
\includegraphics[width=1.0\columnwidth,angle=0,clip=true]{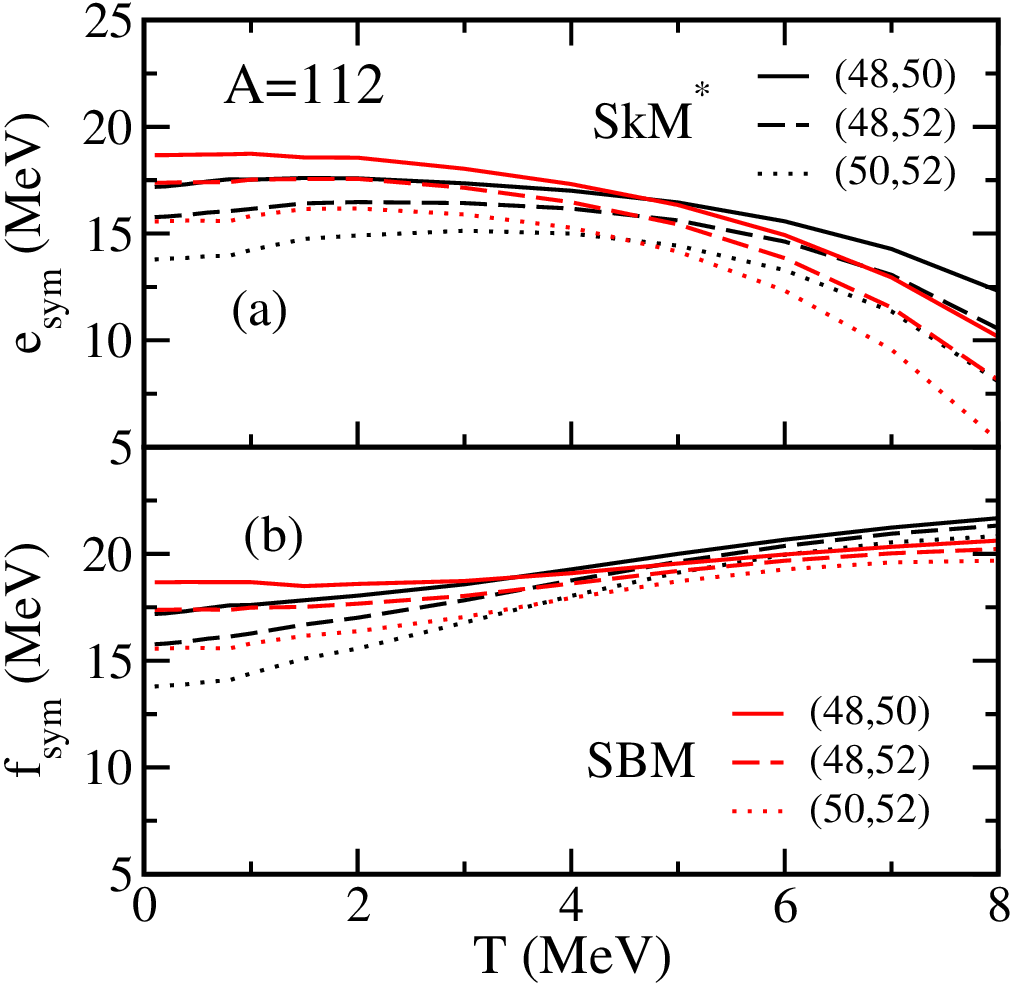}
\caption{(color online) The same as in Fig.~6 for the nucleus A=112.}
\end{figure}

Till now, calculations have been reported without the inclusion of
the effects of the $\omega $-mass. As can be seen from Eq.~(27),
$m_\omega /m$ may be significantly larger than unity, more so for
heavier nuclei and at lower temperatures. This would decrease the
effective kinetic energy of the system and may also decrease the
corresponding symmetry energy. The effects of $\omega $-mass on the
symmetry coefficients $e_{sym}$ and $f_{sym}$ are shown for both the SBM
(red lines)
and SkM$^*$ (black lines) interactions in Fig.~5. 
The number (1) in parentheses refers to calculations (full lines)
with inclusion of the effects of $\omega $-mass, (0) refers to
calculations (dashed lines) without it.
The results from the
difference method are only reported. A relatively light ($A=$56) and 
a medium-heavy ($A=$112) nucleus are chosen for this purpose. As seen
from the upper and lower panels in the figure, $e_{sym}$ and $f_{sym}$
decrease with the inclusion of $\omega $-mass effects. The dilution 
of the effects due to energy mass with increase in temperature is 
evident from the merging of the full and dashed lines with the same
color representing a given interaction at high $T$. As stated earlier,
the reduction in the symmetry coefficients with inclusion of
$\omega $-mass is more prominent for heavier nuclei.

The dependence of the nuclear symmetry energies on the choice of nuclear
pairs for a given isobar is shown in Fig.~6 for $A=$ 56. The chosen
nuclear pairs ($Z_1, Z_2$) are (24,26), (24,28), and (26,28). The
upper panel displays the temperature dependence of $e_{sym}$ and the
lower panel shows the same for $f_{sym}$. The figure shows the sensitivity
of the symmetry coefficients on the choice of the nuclear pair; however,
for all the cases, $e_{sym}$ decreases whereas $f_{sym}$ increases with
temperature. With the nuclear pair $Z_1=$ 26 and $Z_2=$ 28, $e_{sym}$
even becomes negative at high temperature. 
This violates the general understanding of the symmetry energy;
this negativity is likely to arise from the Coulomb polarization
of the density profiles of nuclei. We have checked that calculations
with Coulomb switched off never give negative symmetry energy coefficients.
The different interactions
SkM$^*$ and SBM yield somewhat different values for $e_{sym}$ and
$f_{sym}$; in general, values obtained for the symmetry coefficients
with the SkM$^*$ interaction are lower compared to those for SBM 
interaction at lower temperatures. A crossover, however, is generally
seen at higher temperatures. Calculations have been performed for
nuclear isobars with $A=$ 64 and $A=$ 112 also. They are reported in
Fig.~7 and Fig.~8. These results follow nearly the same trend as
those for $A=$ 56.

\section{Concluding remarks}

 We have performed calculations for the temperature
dependence of the symmetry energy and symmetry free energy coefficients
of atomic nuclei in a Thomas-Fermi model. Calculations have been done
with two effective interactions, namely, the SkM$^*$ and the SBM.
The subtraction procedure has been employed for modeling the hot
nucleus. Effects of the coupling of the surface phonons to the
intrinsic particle motion have been taken into account through a
phenomenological parametric form of the nucleon $\omega $-mass.
The $\omega $-mass increases the nucleon effective mass at relatively
low temperatures, this has an appreciable effect on the symmetry
coefficients which decrease noticeably once effects 
due to $m_\omega$ are taken into account. 

Questions may arise about the proper definition of the symmetry 
coefficients for finite nuclei. We have taken two definitions,
one derived from the usual difference method \cite{dea}, another
from the local density approximation (LDA). In both approaches,
we find that $e_{sym}$ decreases with temperature, $f_{sym}$
shows the opposite trend. In the LDA, the fall in $e_{sym}$
is slow, in the difference method, this fall with temperature is
much stronger. In this method, the results, however, depend 
sensitively on the choice of the nuclear pairs taken to evaluate
the symmetry energies. Occasionally, the symmetry coefficients 
may even be negative, this is violative of the general concept of
the symmetry coefficients.

The ambiguities arising from the different definitions for symmetry
coefficients of finite nuclear systems may possibly be overcome
by calculating the ground state energies of a host of nuclei over
the periodic table with a suitable choice of interaction in a 
microscopic framework, such as Hartree-Fock (HF) and then finding
the mass parameters $a_v, a_s $ {\it etc.} in the spirit of the
liquid-drop model. The calculations can be extended to various
temperatures employing finite temperature HF method and then find the
temperature dependence of the mass parameters which is likely to
give ambiguity-free temperature dependence of the symmetry coefficients.
This is, however, an ambitious program and the work is in progress.

\begin{acknowledgments}
 J.N.D and S.K.S acknowledge support of DST, Government of India.
\end{acknowledgments}

\end{document}